\documentclass[12pt]{iopart}
\usepackage{iopams}
\usepackage{amssymb}
\usepackage{graphicx}

\begin{document}

\newcommand{\ket}[1]{\vert{#1}\rangle}

\title{Statistical properties of the spectrum of the extended Bose-Hubbard model}

\author{Corinna~Kollath}
\address{Centre de Physique Th\'eorique, CNRS, \'Ecole Polytechnique, 91128 Palaiseau Cedex, France}
\ead{kollath@cpht.polytechnique.fr}
\author{Guillaume Roux}
\address{Univ Paris-Sud, Laboratoire de Physique Th\'eorique et Mod\`eles Statistiques, UMR8626, Orsay F-91405, France;
CNRS, Orsay, F-91405, France.}
\author{Giulio Biroli}
\address{Institut de Physique Th{\'e}orique, CEA/DSM/IPhT-CNRS/URA 2306 CEA-Saclay,
F-91191 Gif-sur-Yvette, France}
\author{Andreas M. L\"auchli}
\address{Max Planck Institut f\"ur Physik komplexer Systeme, D-01187 Dresden, Germany}

\date{\today}

\begin{abstract}
  Motivated by the role that spectral properties play for the
  dynamical evolution of a quantum many-body system, we investigate
  the level spacing statistics of the extended Bose-Hubbard model. In
  particular, we focus on the distribution of the ratio of adjacent
  level spacings, useful at large interaction, to distinguish between
  chaotic and non-chaotic regimes. After revisiting the bare
  Bose-Hubbard model, we study the effect of two different
  perturbations: next-nearest neighbor hopping and nearest-neighbor
  interaction. The system size dependence is investigated together
  with the effect of the proximity to integrable points or
  lines. Lastly, we discuss the consequences of a cutoff in the number
  of onsite bosons onto the level statistics.
\end{abstract}


\maketitle

\section{Introduction} 

In classical systems, the chaotic nature of a Hamiltonian plays a
crucial role in determining the dynamics. Chaotic systems explore a
large area of their phase space, whereas non chaotic ones can be
trapped in certain atypical subspaces \cite{Haake2000}. Thus, chaos
plays a considerable role in explaining thermalization and
ergodicity. For quantum systems the situation is less clear. In
principle, the quantum dynamics is linear, since it follows
Schr\"odinger equation, and therefore the notion of chaos is not so
well defined. However, in quantum systems for which a classical
counterpart exists, one finds that the dynamics is very different
depending on whether the corresponding classical motion is chaotic or
regular \cite{Haake2000}.  For generic quantum systems, chaos is also
believed to be essential for thermalization \cite{Peres1984,
  Peres1984a} and delocalization in Fock space \cite{Kota2001}. The
recent realization of closed quantum systems out of equilibrium by
strongly correlated cold atoms clouds \cite{Bloch2008} have lead to a
renewal of interest in the chaotic properties of many-body quantum
systems and their relation to thermalization
\cite{KinoshitaWeiss2006}.

For chaotic quantum systems it has been conjectured that their spectra
show universal features which are related to the theory of random
matrices \cite{Brody1981, Bohigas1986, Mehta1991, Guhr1998}. To
quantify the spectral properties of a given Hamiltonian, a natural
quantity to look at is the gap between adjacent many-body levels
$\delta_n = E_{n+1}-E_n$, where $\{E_n\}$ is the list of eigenvalues
in ascending order. The general symmetries of the Hamiltonian, like
time-reversal and half-integer spin rotational invariance, determine
the random matrix ensemble (among orthogonal, unitary and symplectic
ensemble) to which it belongs~\cite{Brody1981}. A spin-less
time-reversal invariant Hamiltonian such as the Bose-Hubbard model
(for generic parameters) should have similar universal features as the
Gaussian orthogonal ensemble (GOE). For example the adjacent
level-spacing distribution $P_\Delta (\delta)$ is predicted to take
the Wigner-Dyson form
\begin{equation}
P_\Delta (\delta)= \frac{\pi}{2} \frac{\delta}{\Delta}
\exp\left(-\frac{\pi}{4} \frac{\delta^2}{\Delta^2}\right)\;,
\end{equation}
where $\Delta$ is the mean level spacing. On the contrary for
so-called integrable models, in which the properties of the system are
determined by an extensive number of conserved quantities, the
level-spacings should exhibit the following Poissonian distribution
\begin{equation}
P_\Delta (\delta)=  \exp\left(-\delta/\Delta\right)\;.
\end{equation}

The relation between the spectral properties of a quantum system and
the random matrix ensembles has been demonstrated numerically even for
strongly correlated many body systems without classical
counterpart. In particular a GOE like behavior was pointed out for the
non-integrable two dimensional t-J model \cite{Montambaux1993}.  In
one dimensional models, it was checked that at and close to the
integrable points, the statistics was
Poissonian-like~\cite{Poilblanc1993}. In one dimension,
non-diffractive models are integrable on rigorous grounds and solvable
using the Bethe-ansatz which provides all eigenstates
\cite{Sutherland2004}. Several other systems have been discussed since
then \cite{Hsu1993, Guhr1998, Prosen1999, Haake2000}.  Therefore,
analyzing the spectral properties of a system provides a
phenomenological approach to investigate the chaotic nature of the
quantum Hamiltonian.

In this work we consider the properties of the spectrum of the
Bose-Hubbard model. This model is a paradigmatic strongly
correlated many-body system where interactions of amplitude $U$
compete with the kinetic energy favored by the hopping $J$. It
appeared in several contexts of condensed matter theory and regained a
lot of interest since it has been realized in quantum gases confined
to artificial lattice structures \cite{Bloch2008}. Its
out-of-equilibrium dynamics and in particular the question of
thermalization following a quantum quench have also been studied
numerically \cite{Kollath2007, Lauchli2008, Roux2009, Roux2010,
  Biroli2009}.  As perturbations to the Bose-Hubbard Hamiltonian are
experimentally relevant in different regimes \cite{Jaksch1998}, it is
essential to address the sensitivity of the spectral features to extra
terms. We consider the effect of two different perturbations and their
consequences on the level statistics. We focus on the model with a
next-nearest-neighbor hopping with amplitude $J_2$ as well as a
nearest-neighbor interaction with amplitude $V$. The Hamiltonian under study
then reads:
\begin{eqnarray}
\label{eq:bh}
H=&& - J \sum_{j=1}^L \left(b_j^\dagger b^{\phantom{\dagger}}_{j+1}+h.c.\right)- J_2 \sum_{j=1}^L \left(b_j^\dagger b^{\phantom{\dagger}}_{j+2}+h.c.\right) \nonumber \\
&& + \frac{U}{2} \sum _{j=1}^L \hat{n}_j ( \hat{n}_j-1) + V \sum _{j=1}^L \hat{n}_j \hat{n}_{j+1}\;,
\end{eqnarray}
where $b^\dagger_j$ and $b_j$ are the bosonic creation and
annihilation operators, and $ \hat{n}_j = b^\dagger_j
b^{\phantom{\dagger}}_j$ the number operators on site $j$ and $L$ the
number of sites in the chain. Periodic boundary conditions are used to
take advantage of translational invariance. We analyze the spectrum in
the subspace spanned by reflection symmetric states with zero total
momentum using full exact diagonalization. No cutoff in the onsite
occupation $M$ is assumed, i.e. $M=N$, and unit filling $N/L=1$ is
taken if not stated otherwise ($N$ is the total number of bosons). The
symmetrized Hilbert space has dimensions up to $56,822$ for
$N=L=12$. The considered extended Bose-Hubbard model is integrable (in
a broad sense) only in the two limiting cases of $U=V=0$ (free bosons)
and $J=J_2=0$ (atomic limit).

In a first analysis on small lattices, Kolovsky and Buchleitner
\cite{Kolovsky2004} discussed the chaotic behavior of the bare
Bose-Hubbard model ($V=J_2=0$). They investigated the level statistics
and the Shannon entropy of the eigenstates with respect to the basis
of the two integrable limits as a probe of the eigenstates
delocalization (another chaotic feature). A pronounced similarity of
the level statistics with GOE and a maximal Shannon entropy has been
found when the hopping amplitude and the interaction strength are of
the same order. Furthermore, it has been noticed that the low and very
high energy parts of the spectrum seem to display features of
integrable spectra. A chaotic trimeric version of the Bose-Hubbard
model has also been analyzed~\cite{Bodyfelt2007, Hiller2009} and a
comparison to the semi-classical approximation has been carried out
when $N/L \gg 1$.  Another case that has been studied corresponds to
restricting the maximum onsite number of bosons $M$ to one. In this
case, the Bose-Hubbard model boils down to a hard-core boson model
(XXZ) which has the particularity of being integrable provided
$J_2=0$. The level statistics of this hard-core model, perturbed by a
non-integrable operator, has been studied recently
\cite{Santos2010}. While the two models are physically equivalent in
the low-energy physics provided that $U$ is large enough, the
high-energy spectra are very different even in the large-$U$ limit.
  
In this paper, we analyze the level statistics of the extended
Bose-Hubbard model in detail, using system sizes up to $L=N=12$
sites. All data are restricted to the zero-momentum and reflection 
symmetric sector (the one of the ground-state).
In section \ref{sec:bareBH} we study both the level spacings
distribution and the distribution of the ratio of consecutive level spacings. The latter
is a very useful observable, in particular close to the atomic limit.  
This allows us to address the energy and finite size 
dependences. Furthermore, we investigate in section \ref{sec:pertBH}
the influence of different perturbations on the spectrum of the
Bose-Hubbard model. We consider a next-nearest neighbor hopping and a
nearest neighbor interaction and focus
is put on how the chaotic features evolve away from integrable points
(lines) in the parameter space. Lastly, we discuss the effect of the
onsite occupation cutoff $M$ on the level statistics.

\begin{figure}[t]
\centering
\includegraphics[width=0.98\linewidth,clip]{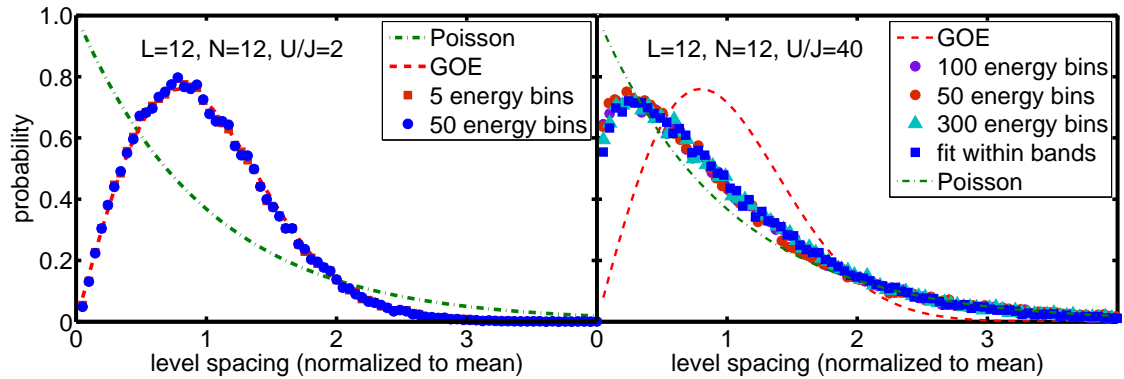}
\caption{The level spacings distribution for $U/J=2$ and $U/J=40$ for a system of length $L=12$ are
  shown (symbols). The dash-dotted, green curve is the Poisson distribution and the dashed, red curve the Wigner Dyson distribution of a purely GOE like ensemble. The different symbols correspond to different unfolding procedures. For the curves labeled by the number of energy bins, the unfolding is performed locally over energy regions, which divides the total energy range covered by the eigenvalues into equally spaced energy bins. The smooth part of the level staircase is fitted by a polynomial. Additionally an unfolding within the energy bands is performed for $U/J=40$ using the function $A2 + (A1-A2)/(1 + \exp((E-E_0)/\textrm{d}E))$ where $A1$, $A2$, $E_0$ and $\textrm{d}E$ are fitting parameters.}
\label{fig:GOE_U}
\end{figure}

\section{Properties of the bare Bose-Hubbard model}
\label{sec:bareBH}

We first focus on the bare Bose-Hubbard model ($J_2=V=0$)
\cite{Kolovsky2004}. In Fig.~\ref{fig:GOE_U},
typical level spacings distributions of the unfolded
spectrum are shown for two different interaction strengths. In order
to remove the dependence on the system specific mean level density,
the 'unfolding' procedure consists in renormalizing the level spacings
by using a suitable fit for the smooth part (see
Ref.~\cite{Guhr1998} for details of the
procedure). The left panel of Fig.~\ref{fig:GOE_U} shows that the distribution for
$U/J=2$ closely follows the Wigner-Dyson distribution of the GOE. In
particular, the distribution vanishes for small level spacings, a
typical signature of level repulsion associated with avoided level
crossings. When approaching the integrable points the distribution
deviates from Wigner-Dyson. This is shown for strong interaction
$U/J=40$ in the right panel of Fig.~\ref{fig:GOE_U}, but occurs also when lowering the
interaction. 
The tail of the distribution is close to the exponential tail of the Poisson distribution;
for small level spacings there is a significant enhancement compared to the GOE
distribution but some level repulsion persists.
We observe that
increasing the size of the system (not shown, see also the discussion here after) tends to increase the
similarity to the Wigner-Dyson distribution. In particular, the repulsion at
small level spacings becomes more and more pronounced. However, for
currently accessible system sizes, it is not clear whether the
distribution very close to the integrable points will converge to the Wigner-Dyson one.

\begin{figure}[t]
\centering
\includegraphics[width=0.5\linewidth,clip]{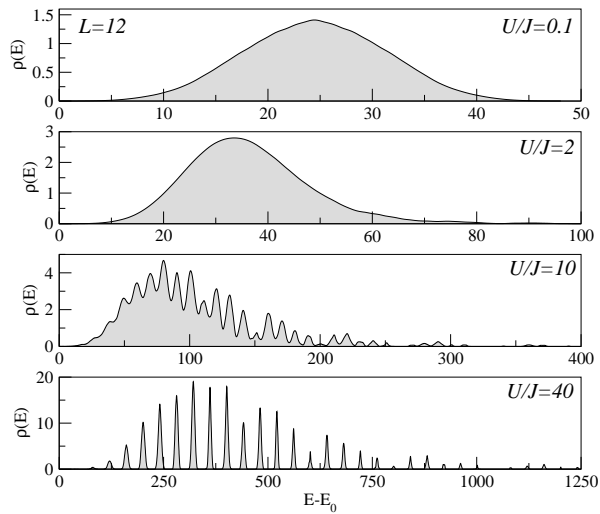}
\caption{The many-body density of states $\rho$ in the zero-momentum and reflection symmetric sector and for different interaction strengths. $E_0$ is the ground-state energy. The evolution from a smooth spectrum at low interaction strength to the appearance of well separated energy bands at large interaction strength is shown. The smooth curves are obtained broadening the $\delta$-peaks by Gaussians ($\bar{\rho}(E)=\sum_n g_{\Delta}(E-E_n)$ with $g_{\Delta}$ a Gaussian of width $\Delta$). However, note that even for large interaction strengths the width of the Gaussians is much smaller than the width of the energy bands.}
\label{fig:DOS}
\end{figure}

The study of systems with large interaction strength is involved
due to the appearance of a band structure in the density of states (see Fig.~\ref{fig:DOS}). The spectrum evolves from a smooth broad spectrum for small interaction strength to a series of narrow energy bands separated by $U$ for large interaction strength\footnote{However, note that the mean level spacing is still much smaller than the width of the energy bands.}. This would make the unfolding complicated because it is difficult to separate  the spectrum into a smooth and a fluctuating part. To obtain reliable results we performed different unfolding procedures fitting locally different parts of the spectrum. If the ranges over which the fits are performed are chosen in a suitable way, the same general form of the spectra is recovered, even though small discrepancies can occur. Note that the complicated form of the density of states also makes a study of longer range correlations of the spectrum involved.   

\begin{figure}[t]
\centering
\includegraphics[width=0.8\linewidth,clip]{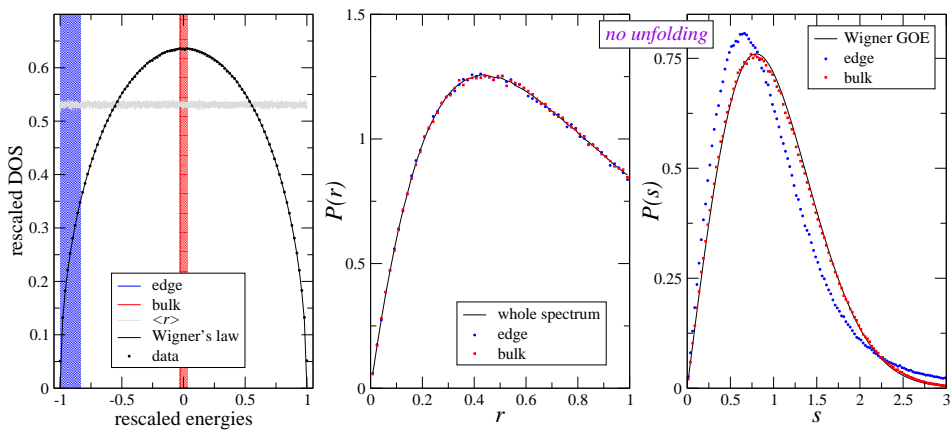}
\caption{Left panel: semicircular density of state for an GOE ensemble of random matrices of size $5,000^2$ averaged over $10,000$ realizations. In addition, the value of the ratio of consecutive level spacings, averaged over the different realizations, is given as a function of energy, showing that it does not depend on the density of states. The shaded regions are the regions over which the distributions shown in the central and right panel are taken (they both contain 200 states).
Central panel: Distribution of the ratio of consecutive level spacings taken over a region at the edge, in the bulk and of the whole spectrum: the distribution does not depend on the variation of the local density of states. Right panel: Distribution of the level spacing in the same regions without unfolding the spectrum showing the strong dependence on $\rho(E)$.}
\label{fig:comp_s_r}
\end{figure}

\begin{figure}[t]
\centering
\includegraphics[width=0.7\linewidth,clip]{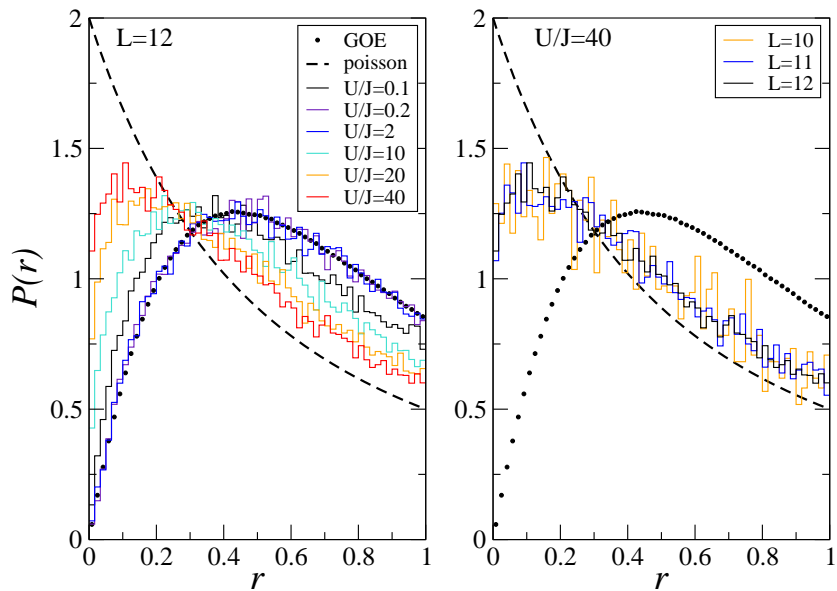}
\caption{Typical distribution of the ratio of adjacent level spacings
  $r$. Left panel: Different interaction strengths for $L=12$ are
  compared with Poisson and GOE distributions. The proximity to the
  chaotic regime is seen when $U\simeq J$. While for $U\gg J$ and $U\ll J$
  the curves approach the Poissonian tail. Right panel: system
  size dependence for $U/J=40$.}
\label{fig:Rdistrib}
\end{figure}

To avoid this complication, we continue our study using another measure which has the
advantage of not depending on the unfolding procedure: the ratio of
consecutive gaps between adjacent levels
\cite{Oganesyan2007}. This quantity is defined by $r_n=\min(\delta_n,
\delta_{n-1})/\max(\delta_n, \delta_{n-1}) $. 
As long as the density of states does not vary on the scale of the mean level spacing, the trivial dependence on the smooth part drops out and there is no need for unfolding. This is exemplified in Fig.~\ref{fig:comp_s_r}, where we compare for a GOE random matrix ensemble  the distribution of the level spacing and of the ratio of the consecutive level spacings taken over two different energy ranges without performing the unfolding procedure. The first energy range lies at the boundary of the spectrum where the density of states varies considerably. In contrast the second energy range is situated in the center of the spectrum, where the density of states exhibits only slow changes. The GOE distribution is computed numerically as in
Ref.~\cite{Oganesyan2007} using the averaged results of $10,000$ samples of
random matrices of size $5,000^2$. It is clearly seen that the ratio of consecutive level spacings does not depend on the region used, whereas the level spacing statistics is different for the two chosen regions due to its strong dependence on the smooth part of the spectrum. The local ratio of consecutive level spacings averaged over the GOE ensemble is also independent of the density of states (left panel of Fig.~\ref{fig:comp_s_r}). 
On the basis of these results, which show that $r$ is a useful random variable to analyze the spectral statistics, we continue our study focusing on the ratio of consecutive gaps. 

We give in the left panel of Fig.~\ref{fig:Rdistrib} typical distributions $P(r)$
obtained for the Bose-Hubbard model at different interaction strength. As expected, the $P(r)$ distribution depends on the chaoticity of the Hamiltonian:
the distribution for a Poissonian spectrum reads $P(r)=2/(1+r)^2$ while it can been computed numerically for the GOE ensemble (Fig.~\ref{fig:comp_s_r} and Fig.~\ref{fig:Rdistrib}). As for the level spacing distribution, the maximum resemblance with GOE is observed when $U \simeq J$. In the proximity of the two integrable limits $U=0$ and $J=0$ the distributions approach the Poisson prediction, particularly
on the tail. We checked for a wide range of values of $U$ that the distribution of these ratios is in good agreement with the level spacing distribution using different local unfolding procedures. 

\begin{figure}
\centering
\includegraphics[width=0.65\linewidth,clip]{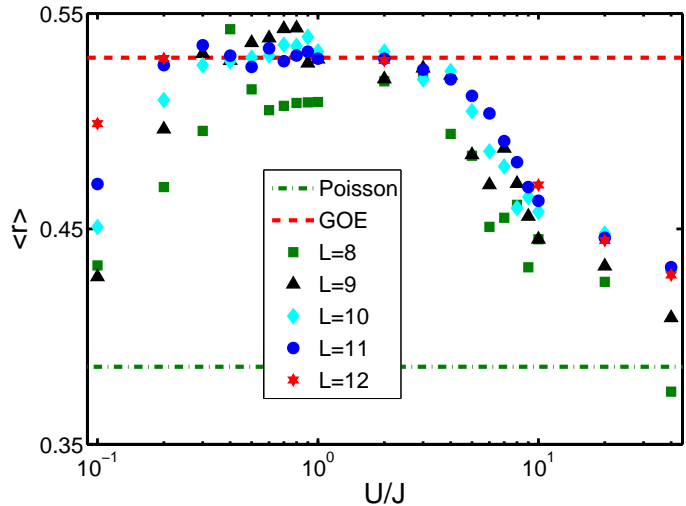}
\caption{Evolution of the average ratio of consecutive level spacings
  $\langle{r}\rangle$ with the interaction strength $U/J$. For
  the Poisson distribution the average value of
  $\langle{r}\rangle_P=0.386$ and for the GOE ensemble
  $\langle{r}\rangle_{GOE}= 0.53$.  The average is taken for the
  full spectrum, without cutting the low and the high energy part. A
  clear tendency of larger system sizes versus the GOE ensemble value
  is seen.}
\label{fig:ratio}
\end{figure}

\begin{figure}
\centering
\includegraphics[width=0.65\linewidth,clip]{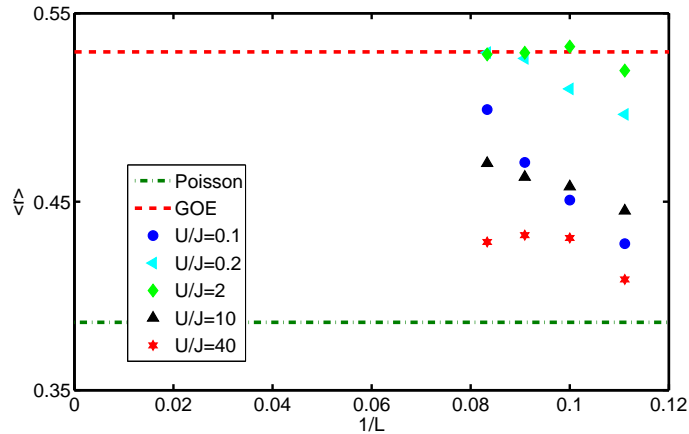}
\caption{Scaling of the ratio of consecutive level spacings
  $\langle{r}\rangle$ with the inverse system size for different interaction
  strength.}
\label{fig:ratioscaling_length}
\end{figure}

In order to have a systematic tool to probe the proximity
either to Poisson or GOE, we use the results on the mean value
$\langle{r}\rangle$ which is $\langle{r}\rangle_P=2\ln 2-1\approx
0.386$ for Poisson and $\langle{r}\rangle_{GOE} \approx 0.53(1)$ for
GOE.  We thus expect that this averaged ratio should display a maximum
for an interaction strength around $U\approx J$. In
Fig.~\ref{fig:ratio} this ratio is given for different interaction
strengths and system lengths. At intermediate interaction strength
($0.3<U/J<4$) we see that $\langle{r}\rangle$ gets very close to the
GOE prediction. Even though there is a small system size dependence
left, the values for the longer system sizes considered are quite
close to the expected value. For small and large values of the
interaction strength we see that the behavior is different. For these
regimes the values of $\langle{r}\rangle$ lie in between the ones
expected for the GOE and the Poisson ensemble and for most of these
values a strong system size dependence is still evident. Typically,
the trend for longer system sizes goes towards
$\langle{r}\rangle_{GOE}$, as shown in
Fig.~\ref{fig:ratioscaling_length} for some chosen values of the
interaction. This suggests that away from the integrable points
some critical length scale (or particle number) exists above which the
system shows a level statistics which is very close to the one of the
GOE ensemble.  Our results further suggest that this length scale
possibly grows in the proximity of the integrable points.  For
instance, when $U/J=40$, the distributions hardly evolve with the
system size (see Fig.~\ref{fig:Rdistrib} right panel). We expect that
for large enough sizes the properties of the spectrum might be well
described by a GOE.
However, the question whether or not a finite deviation of the
parameters from the integrable limit is necessary to obtain GOE like
characteristics cannot be conclusively answered\footnote{Notice that
  the scaling analysis of the level statistics have some strong
  numerical limitations. Indeed, as the width of the spectrum scales
  as $N^2$ while the number of states scales exponentially with $N$,
  we may expect the minimal level spacing to reach the numerical
  accuracy of full diagonalization at some relatively small system
  size $L$. However, these system sizes are longer than the here
  considered system sizes.}.
Still, the obtained results can be compared with other scenarios on the effect of
the interaction in the Bose-Hubbard model. For instance, Cassidy \etal
suggested~\cite{Cassidy2009} that there could be an interaction
threshold in the Bose-Hubbard model for the chaotic behavior to
develop, based on calculations valid in the semi-classical limit $N/L
\gg 1$ supplemented by a mean-field calculation. Extrapolating their
results to the $N=L$ limit, the threshold would be $U/J \simeq
0.5$. In contrast, our results demonstrate that for low filling $n=1$
even at $U/J \simeq 0.1$, the level statistics features have a strong
tendency towards a chaotic behavior and no signature of a threshold is
found.


\begin{figure}[t]
\centering
\includegraphics[width=0.6\linewidth,clip]{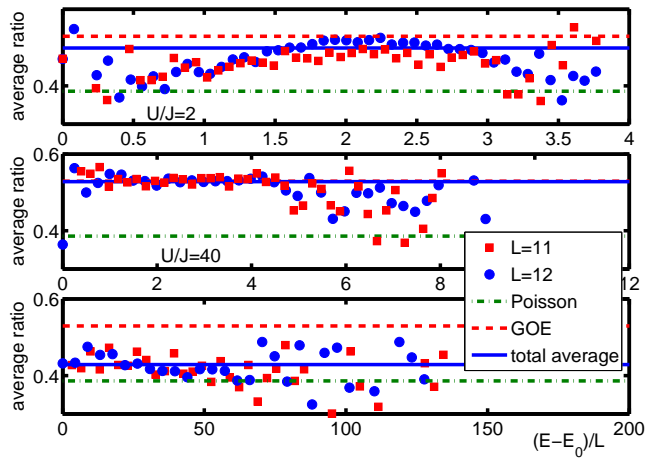}
\caption{
Average ratio of consecutive level spacings taken in different energy intervals of equal size for different interaction strength $U/J=0.1,2, 40$ and different system length $L=11,12$. The number of bins taken for the average is 50.
}
\label{fig:levellocal}
\end{figure}

Up to now we have considered the properties of the whole spectrum. However the question arises how these properties change within different energy ranges of the spectrum. To illustrate this, we show in Fig.~\ref{fig:levellocal} the average ratio taken over different ranges of the spectrum\footnote{We checked that the ratio exhibits the same features as the level spacing statistics.}. We must notice that contrary to the GOE benchmark of Fig.~\ref{fig:comp_s_r}, the Hamiltonian is deterministic and no sampling can smoothen the curves: we were able to get good statistics only by reaching large enough systems sizes. For small values of the interaction strength ($U/J=0.1$ and $U/J=2$ in Fig.~\ref{fig:levellocal}) the average ratio slightly depends on the energy and one can observe that the bulk of the spectrum displays the GOE prediction while edges show some deviations with strong fluctuations. These strong fluctuations may be attributed either to the fact that the small density of states induces bad statistics or to the fact that the physics at the edges (in particular close to the ground-state) display different statistics than for the high energy excited states in the bulk. A clear maximum exists in the central region. Increasing the system size for $U/J=0.1$ makes the bulk value closer to the GOE prediction. For intermediate values of the interaction strength $U/J=2$ (central panel Fig.~\ref{fig:levellocal}) the properties of the spectrum do not change much in different energy regions and are close to GOE. For large values of the interaction strength (lower panel in Fig.~\ref{fig:levellocal}) a band structure develops in the energy spectrum. The ratio shows stronger fluctuations\footnote{The ratio and its fluctuations also depend more strongly on the energy interval used. However we checked that the main trend remains for typical energy ranges taken.} and a slight drop from a more Wigner-Dyson like value towards a more Poisson like value.

\begin{figure}
\centering
\includegraphics[width=0.45\linewidth,clip]{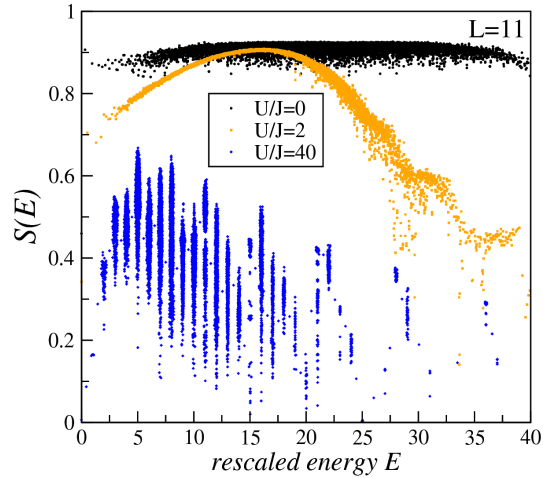}
\caption{The Shannon entropy in the real-space symmetrized basis of exact diagonalization is shown for different interaction strength. An evolution from a smooth behavior for small interaction strength towards a very fluctuating behavior for large interaction strength can be seen.}
\label{fig:entropy}
\end{figure}

\begin{figure}
\centering
\includegraphics[width=0.45\linewidth,clip]{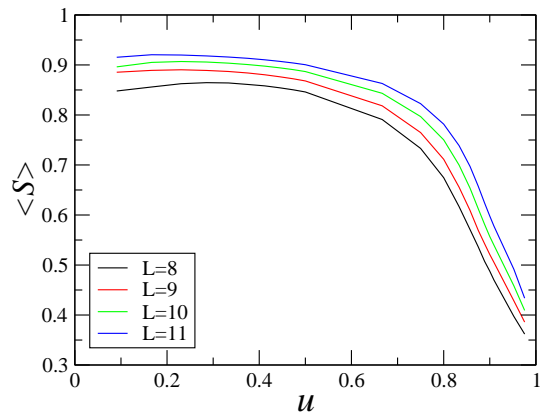}
\caption{The finite system dependence of the average value of the Shannon entropy is shown. A clear trend towards larger values is found.}
\label{fig:averageentropy}
\end{figure}

In addition to the distributions discussed above, the Shannon entropy can give interesting information about the chaoticity of the system  \cite{Kolovsky2004}. The Shannon entropy is defined by $S_n=(-\sum_m \vert{c_m^n}\vert^2\ln\vert{c_m^n}\vert^2)/\ln\textrm{dim}H$, where $c_n^m$ are the coefficients of the $n$th eigenstate decomposed onto the $m$th basis state of a chosen basis. The entropy measures the delocalization of a wave-vector with respect to a chosen basis: with our definition, it reaches 1 for a fully delocalized state. 
Here we measure the entropy with respect to the symmetrized real-space basis of exact diagonalization. In Fig.~\ref{fig:entropy} the behavior of the entropy is shown for different interaction strengths. A clear evolution from a smooth to a sharp distribution around high values at small interaction strength towards a strongly fluctuating distribution at low values is evident. This shows that for small interaction strength almost all eigenvectors are delocalized as expected in the non-interacting regime where the Hamiltonian is diagonal in the momentum space. At intermediate interaction strength, the edges display less delocalized features than in the bulk. The most interesting behavior is exhibited at large interaction strength where the degree of localization fluctuates strongly between different eigenvectors within a Mott lobe. Let us point out that this finding is similar to the behavior of some observables and weights calculated in these eigenstates and that was identified to be the reason of non-thermalization after a quench on such finite size systems \cite{Roux2009, Roux2010,Biroli2009}. Finally, let us comment on the evolution of the above results increasing the system size. We find that even though Fig.~\ref{fig:averageentropy} displays a trend towards delocalization, the behavior in the thermodynamic limit is particularly hard to access close to the infinite-$U$ integrable point.

\section{Perturbing the Bose-Hubbard model} 
\label{sec:pertBH}

\begin{figure}
\centering
\includegraphics[width=0.9\linewidth,clip]{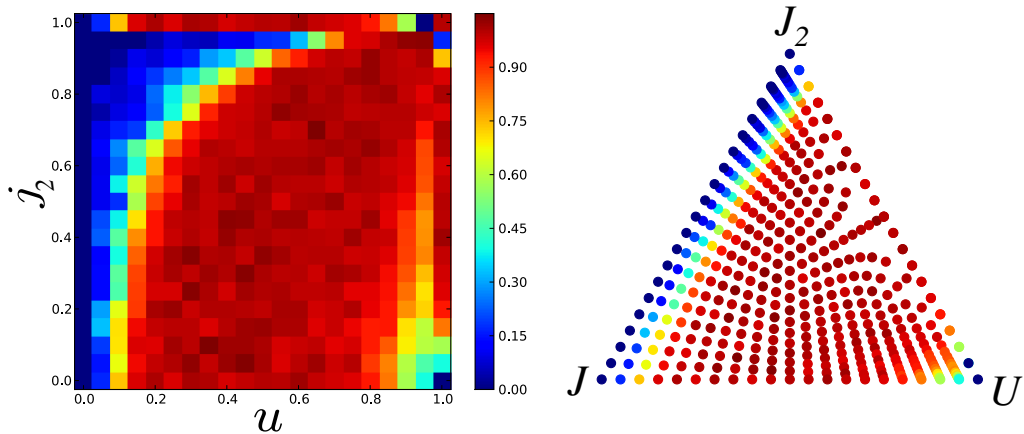}
\caption{Density plots of the evolution with the parameters $j_2$ and
  $u$ (see text) of the averaged ratio of consecutive level spacings
  $\langle{r}\rangle$ ($L=11$). We plot $(\langle{r}\rangle -
  \langle{r}\rangle_P) / (\langle{r}\rangle_P -
  \langle{r}\rangle_{GOE})$, so that blue (resp. red) corresponds to
  Poissonian (GOE) distribution. Right panel: ternary plot of the same
  data showing the integrable lines $J-J_2$ and the isolated
  integrable point ($U$ corner).}
\label{fig:triangle_J2}
\end{figure}

We now turn to the effect of the $J_2$ and $V$ perturbations
(separately) that are expected to help breaking integrability at the
$J=0$ and $U=0$ integrable points, respectively.  As there are three
parameters ranging from zero to infinity, we will fold the parameter
space using two representations (see Fig.~\ref{fig:triangle_J2} for an
example). The first one is to introduce the function $f(x,y)=x/(x+y)$
and to use the following definition: $u=f(U,J)$ and $j_2=f(J_2,J)$
when $J\neq 0$; $u=f(U,J_2)$ and $j_2=f(J_2,U)$ if $J=0$, with the
additional point $u=j_2=1$ when $U=J_2$. Such a folding is useful to
restrict the considered parameters onto a finite interval, and enables
one to easily deduce the parameters for a given point. One
disadvantage of this folding are discontinuities arising from the
infinities on the $j_2=1$ and $u=1$ lines. A continuous way to draw
the data is to use a ternary plot\footnote{Formally speaking, a point
  of the diagram corresponds to percentage of each parameter, i.e. a
  triplet $(\%J,\%J_2,\%U)$. In cartesian coordinates with the triplet
  (100,0,0) at the origin, one has $x = (U+J_2/2)/\mathcal{N}$ and $y
  = \sqrt{3} J_2 / 2\mathcal{N}$ with $\mathcal{N} = J + J_2 +
  U$.}. However, it is more difficult to find in such a plot the
original parameters. Therefore, we use both ways to present our
results.
 
\begin{figure}
\centering
\includegraphics[width=0.6\linewidth,clip]{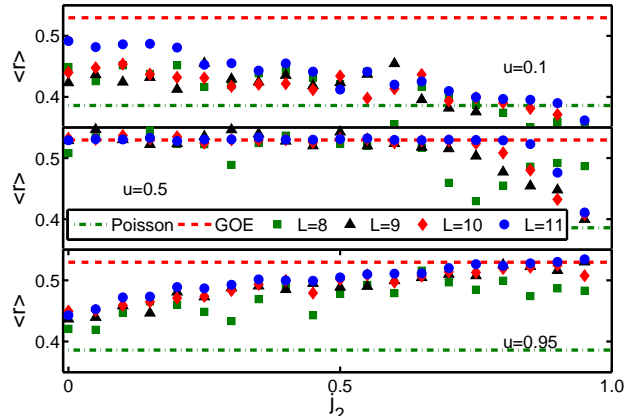}
\caption{Evolution of the ratio of consecutive level spacings
  $\langle{r}\rangle$ with nn hopping $j_2$ for different interaction
  strengths. For the Poisson distribution the average value of
  $\langle{r}\rangle_P=0.386$ and for the GOE ensemble
  $\langle{r}\rangle_{GOE}= 0.53$ are displayed.  The average is
  taken for the full spectrum, without cutting the low and the high
  energy part (circles).}
\label{fig:ratio_J2_Ucomb}
\end{figure}

\begin{figure}
\centering
\includegraphics[width=0.45\linewidth,clip]{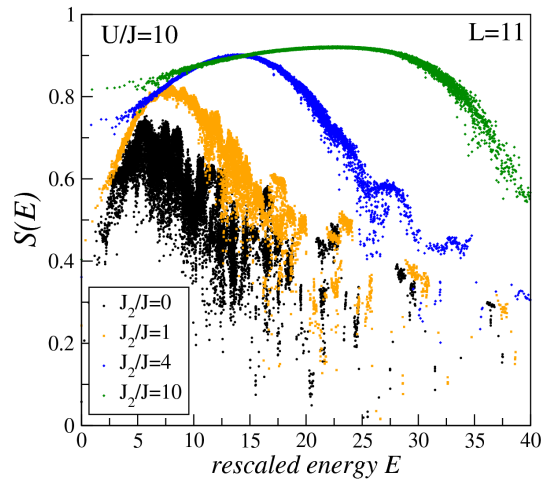}
\caption{Effect of the $J_2$ perturbation on the Shannon entropy in the symmetrized real-space basis.}
\label{fig:entropy_J2}
\end{figure}

\paragraph{Influence of a next-nearest neighbor hopping} -- If the
next-nearest neighbor hopping $J_2$ is switched on, the behavior of
the spectrum changes. A summary of the effect of $J_2$ for $L=11$ is
presented in Fig.~\ref{fig:triangle_J2}. For small and intermediate
interaction strength, the additional finite value of $J_2$ drives the
small systems closer to the Poisson behavior. For large interaction
strength, $J_2$ helps to drive the system away of the integrable point
$J=0$, before at very large $j_2$ it again turns Poisson like due to
the attraction of the $U=J=0$ integrable point. The values of $J_2$
can thereby be much smaller than the actual interaction strength and
still have a considerable influence. In the ternary diagram the almost
symmetric behavior of the system with respect to the diagonal $J=J_2$
is nicely visible. This means that the next-nearest neighbor hopping
$J_2$ has a similar effect than the nearest neighbor hopping $J$.  In
order to discuss the influence of longer system sizes, we show in
Fig.~\ref{fig:ratio_J2_Ucomb} the dependence of the average ratio
$\langle{r}\rangle$ on $j_2$ for different system sizes at chosen
values of the interaction strength. If the behavior is already close
to the GOE one, finite size effects are typically very small
(Fig.~\ref{fig:ratio_J2_Ucomb} central panel). In contrast, if the
system is not in the GOE regime, the finite size effects become more
pronounced. However, the larger system sizes show a clear tendency
towards the GOE behavior. Lastly, we have checked that the same
qualitative features are displayed in the Shannon entropy when $J_2$
is taken into account. For instance, Fig.~\ref{fig:entropy_J2} shows
that starting from a large $U/J$ and increasing $J_2$ tends to make the
wave-functions delocalize in the symmetrized real-space basis of exact
diagonalization. In particular, for the point with $J=J2$, the perturbation
clearly favors delocalization. When $J_2/U$ increases, the delocalization with respect to the real space 
basis becomes more and more pronounced due to the dominating kinetic 
term. As for the case of the pure Bose-Hubbard model, a strong 
dependence on the energy is observed.
A maximum in the Shannon entropy is found for intermediate energies, 
whereas for low and in particular high energy the Shannon entropy drops 
quickly and exhibits typically more fluctuations.

\begin{figure}
\centering
\includegraphics[width=0.9\linewidth,clip]{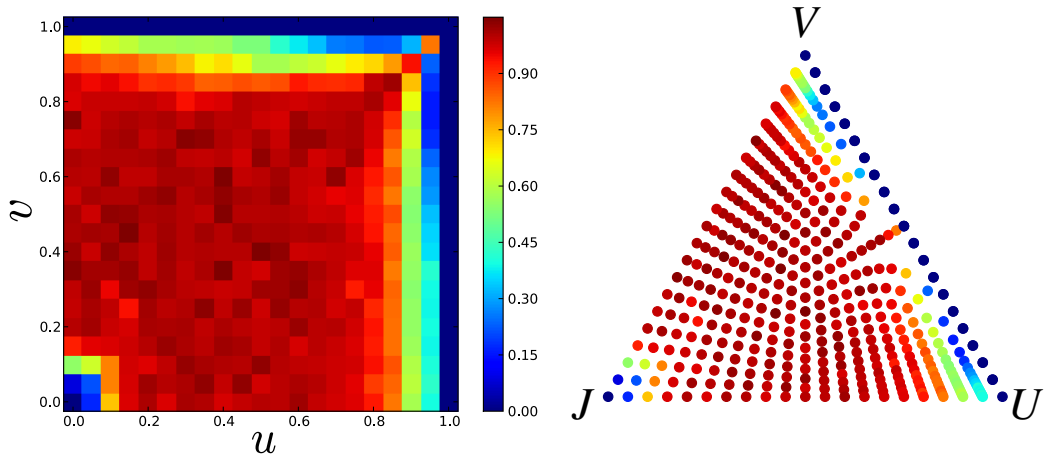}
\caption{Same as Fig.~\ref{fig:triangle_J2} changing $J_2$ with
  nearest-neighbor interaction $V$.}
\label{fig:triangle_V}
\end{figure}

\paragraph{Influence of the nearest-neighbor interaction} -- In the
following paragraph we discuss the influence of nearest-neighbor
interactions which is summarized in Fig.~\ref{fig:triangle_V}. The
same representations of the parameters space are taken, replacing
$J_2$ with $V$ and $j_2$ with $v$. The behavior is qualitatively very
different from the $J_2$ perturbation: the integrable line in the
ternary plot is now the $U-V$ line. In both representations, the data
are nearly symmetrical with respect to the diagonal $U=V$. Thus, the
two interacting terms act in a similar way in terms of level
statistics. For small onsite interaction strength $U$, a finite
nearest neighbor interaction $V$ enhances the trend of the ratio
towards its GOE value.  However--surprisingly at first sight--at large
onsite interaction strength $U$, a small finite value of $V$ induces a
trend towards the GOE like behavior and only if $V$ is larger than the
onsite interaction $U$ the value of the ratio drops drastically to the
Poisson value. In particular for interactions of the same order of
magnitude $U \sim V$, $J$ rapidly drives the system towards GOE. This
effect can be made plausible in a simplified picture considering the
eigenstates of the Hamiltonian at $J=0$, which are Fock
states. However, their order with respect to the energies is very
different for both interactions. To make this more explicit consider
the state with one particle per site $\ket{1}=\ket{1,1,\dots,1}$ and
the state with two particles every second site
$\ket{2}=\ket{2,0,2,0,\dots,2,0}$. For a strong onsite interaction the
state $\ket{1}$ is very low in energy whereas the state $\ket{2}$ lies
in the upper part of the spectrum. In contrast, for a strong
nn-interaction state $\ket{2}$ lies in the lower part of the spectrum
whereas state $\ket{1}$ lies in the upper part. If both interactions
are of the same order of magnitude both states lie very close in
energy such that the small hopping has a large effect on the
states. These two energy states are examples of the behavior of many
of the energy states which become almost degenerate in the limit of
equal onsite and nn-interaction strength. Therefore, the effect of the
hopping as a perturbation is expected to be more effective when $U\sim
V$ and should help make the level statistics GOE-like.

\begin{figure}
\centering
\includegraphics[width=0.6\linewidth,clip]{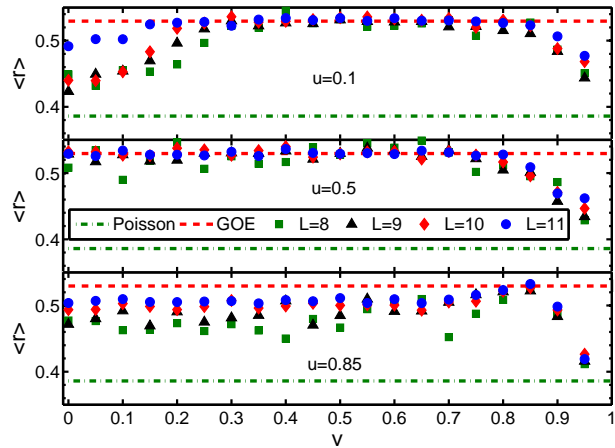}
\caption{Same as Fig.~\ref{fig:ratio_J2_Ucomb} changing $J_2$ with
  nearest-neighbor interaction $V$.}
\label{fig:ratio_V_Ucomb}
\end{figure}

In Fig.~\ref{fig:ratio_V_Ucomb}, finite size effects are considered.
As for the other discussed cases the finite size effects are very
small if the value of the ratio is already close to GOE. For the
remaining values, we typically see a trend of the ratio for larger
system sizes towards GOE.

\begin{figure}
\centering
\includegraphics[width=0.6\linewidth,clip]{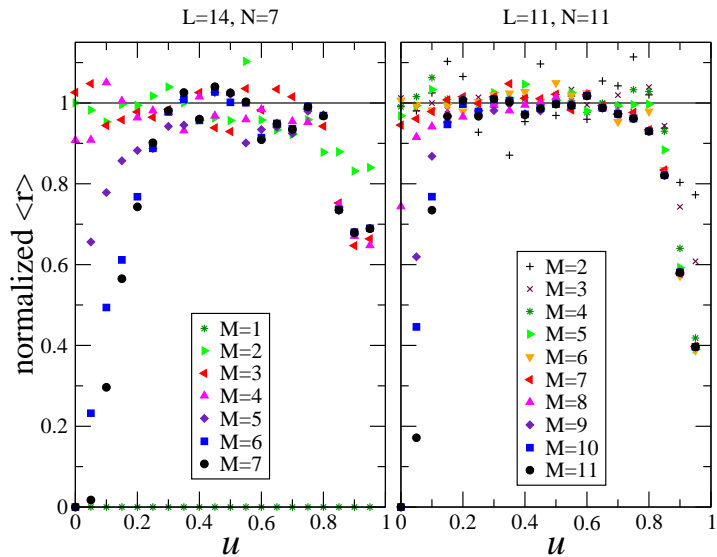}
\caption{Evolution of $(\langle{r}\rangle - \langle{r}\rangle_P) /
  (\langle{r}\rangle_P - \langle{r}\rangle_{GOE})$ with the maximum
  number of onsite bosons $M$. The average is taken for the full
  spectrum, without cutting the low and the high energy part
  (circles). Left panel: incommensurate density where hard-core bosons
  corresponds to $M=1$. Right panel: the same at a commensurate
  density $N=L$.}
\label{fig:ratio_cutoff}
\end{figure}

\paragraph{Occupation cutoff dependence} -- In
Fig.~\ref{fig:ratio_cutoff}, we show the effect of introducing a
cutoff in the number of bosons $M$ on each site. For large values of
the interaction strength the use of a cutoff $M>4$ does not seem to
change the spectral properties much, both at unit and half fillings. A
smaller cutoff strongly affects the ratio, pushing it close to the GOE
prediction. This is due to the suppression of the remaining particle
fluctuations mixing the true eigenstates which are no longer
represented. For intermediate interaction where the ratio is close to
the GOE value, the effect of the cutoff on the mean ratio is
relatively small. In contrast for small interaction strength the
influence of the cutoff is very pronounced. Here the introduction of a
small cutoff does drive the system away from integrability. Even for
$U=0$, using $M<N$ makes the levels statistics close to GOE. This
effect can be understood by recalling the properties of the
eigenstates in the limit of weak interaction. These are the momentum
eigenstates which comprise strong particle fluctuations. If one
introduces a cutoff for the number of bosons per site these states
cannot be represented anymore and start to mix. In other words: the
local constraint is equivalent to using a projector on the kinetic
part which correlates the bosons or, equivalently, acts as a
complicated effective interaction that turns out to display a GOE
behavior. A qualitatively similar effect is found in the comparison
between the 1D t-J and Hubbard model. The Hubbard model is integrable
while the t-J model which is related to the Hubbard model by
Gutzwiller's projection is generically not~\cite{Poilblanc1993}.

In addition, we can discuss earlier results in the literature
addressing the question of the integrability of the 1D Bose-Hubbard
model and the effect of multi-occupancies. Seminal studies by Choy and
Haldane~\cite{Choy1980, Haldane1980} seemed to argue that Bethe-ansatz
equations yield solutions of Bose-Hubbard-like models but the analysis
turned out to be invalid~\cite{Haldane1981, Choy1982}. The authors
emphasized that $M>3$ was required to give rise to non-integrability.
Later, Krauth~\cite{Krauth1991} used the Bethe-ansatz wave-function as
a variational approach for the ground state properties. He found that
for $U/J \lesssim 2$, the comparison with unbiased quantum Monte-Carlo
results was indeed very good. This finding supported the fact that the
integrable nature of the free bosons gas was preserved up to
interactions close to the transition point to the Mott insulating
phase at least for ground state properties. The results of the present
study, in which we found considering the entire spectrum that the
chaotic properties emerge much below $U/J\simeq 2$ ($u\simeq 0.66$),
stress the difference between the low-energy part of the spectrum and
high-energy regions: the ground-state and first excitations might have
integrable-like behavior (if the density of quasi-particles is small,
they may interact less) while one cannot consider a high energy
excitation as simply being made of a superposition of elementary
excitations~\cite{Montambaux1993} (a picture which survives high in
energy in the Bethe-ansatz and in free particles systems).

\section{Conclusion}

To conclude, we presented a study of the characteristic properties of
the spectra of the extended Bose-Hubbard model. In an intermediate
regime of the interaction strength the system is in the GOE regime. In
contrast for very weak and strong interaction strength the analysis
suggests an approach toward GOE when increasing system sizes. In most parameter regimes this trend towards the GOE was most pronounced in the central region of the spectrum. An
additional next-neighbor hopping amplitude $J_2$ changes the
properties of the energy levels in these small systems. It acts
similar to the hopping amplitude $J$. For weak interaction, $J_2$
drives the system closer to the Poisson like behavior, whereas for
large interaction strength it reinforces the GOE like behavior.  An
additional nearest neighbor interaction $V$ has a similar effect on
the statistical properties of the spectrum as the onsite interaction
even though the corresponding eigenstates are very differently
distributed in energy. Close to the point where the interaction $U$
and $V$ become of similar strength even a very small value of $J$ is
enough to induce a GOE like statistics.  Finally we discussed the
influence of the introduction of a cutoff for the number of bosons per
site usually used to render the system numerically tractable. Here we
see that the cutoff can change the statistics of the spectrum from
Poisson like to GOE like, in particular at small interaction strength.

We see that for all the different regimes considered the changes with
increasing system size can be divided into two main classes. If the
properties of the system are already GOE like, increasing the system
size only induces small changes. In contrast if the value of ratio of
consecutive level spacings lies in between the Poisson and the GOE
value indicating a mixed statistics, finite size effect are
considerable. In this regime larger system sizes typically drive the
system towards the GOE value indicating a GOE like behavior in the
thermodynamic limit. However, larger sizes would be needed to obtain a
conclusive result on the question whether there is always a large
enough system size to reach a GOE behavior for all parameters except
the ones corresponding to the integrable points (or lines) or if a
threshold for the perturbation from the integrable point exists to
reach it.

\ack

We would like to thank B. Altshuler, N. Andrei, and A. Millis for
fruitful discussions. This work was partly supported by the 'Triangle
de la Physique', DARPA-OLE, and the ANR ('FAMOUS').

\section*{References}
\bibliographystyle{iopart-num}
\bibliography{ref}

\providecommand{\newblock}{}
\begin{thebibliography}{10}
\expandafter\ifx\csname url\endcsname\relax
  \def\url#1{{\tt #1}}\fi
\expandafter\ifx\csname urlprefix\endcsname\relax\def\urlprefix{URL }\fi
\providecommand{\eprint}[2][]{\url{#2}}

\bibitem{Haake2000}
Haake F 2000 {\em Quantum Signatures of Chaos\/} (Springer, Berlin Heidelberg
  New York)

\bibitem{Peres1984}
Peres A 1984 {\em Phys. Rev. A\/} {\bf 30} 1610--1615

\bibitem{Peres1984a}
Peres A 1984 {\em Phys. Rev. A\/} {\bf 30} 504--508

\bibitem{Kota2001}
Kota V~K~B 2001 {\em Phys. Rep.\/} {\bf 347} 223

\bibitem{Bloch2008}
Bloch I, Dalibard J and Zwerger W 2008 {\em Rev. Mod. Phys.\/} {\bf 80} 885

\bibitem{KinoshitaWeiss2006}
Kinoshita T, Wenger T and Weiss D~S 2006 {\em Nature\/} {\bf 440} 900

\bibitem{Brody1981}
Brody T~A, Flores J, French J~B, Mello P~A, Pandey A and Wong S~S~M 1981 {\em
  Rev. Mod. Phys.\/} {\bf 53} 385--479

\bibitem{Bohigas1986}
Bohigas O and Giannoni M 1986 {\em Quantum Chaos and Satistical Nuclear
  Physics\/} ({\em Lect. notes Phys.\/} vol 263) (Springer (Berlin))

\bibitem{Mehta1991}
Mehta M~L 1991 {\em Random Matrices\/} 2nd ed (Academic Press)

\bibitem{Guhr1998}
Guhr T, Mueller-Groeling A and Weidenmueller H~A 1998 {\em Phys. Rep.\/} {\bf
  299} 189--425

\bibitem{Montambaux1993}
Montambaux G, Poilblanc D, Bellissard J and Sire C 1993 {\em Phys. Rev.
  Lett.\/} {\bf 70} 497--500

\bibitem{Poilblanc1993}
Poilblanc D, Ziman T, Bellissard J, Mila F and Montambaux G 1993 {\em Europhys.
  Lett.\/} {\bf 22} 537--542

\bibitem{Sutherland2004}
Sutherland B 2004 {\em Beautiful Models\/} (Singapure: World Scientific)

\bibitem{Hsu1993}
Hsu T~C and Angl\`es~d'Auriac J~C 1993 {\em Phys. Rev. B\/} {\bf 47}
  14291--14296

\bibitem{Prosen1999}
Prosen T 1999 {\em Phys. Rev. E\/} {\bf 60} 3949--3968

\bibitem{Kollath2007}
Kollath C, L\"{a}uchli A~M and Altman E 2007 {\em Phys. Rev. Lett.\/} {\bf 98}
  180601

\bibitem{Lauchli2008}
L\"auchli A~M and Kollath C 2008 {\em J. Stat. Mech.\/}  P05018

\bibitem{Roux2009}
Roux G 2009 {\em Phys. Rev. A\/} {\bf 79} 021608

\bibitem{Roux2010}
Roux G 2010 {\em Phys. Rev. A\/} {\bf 81} 053604

\bibitem{Biroli2009}
Biroli G, Kollath C and L\"auchli A 2009  (\textit{Preprint}
  \eprint{arXiv:0907.3731})

\bibitem{Jaksch1998}
Jaksch D, Bruder C, Cirac J~I, Gardiner C~W and Zoller P 1998 {\em Phys. Rev.
  Lett.\/} {\bf 81} 3108--3111

\bibitem{Kolovsky2004}
Kolovsky A~R and Buchleitner A 2004 {\em Europhys. Lett.\/} {\bf 68} 632--638

\bibitem{Bodyfelt2007}
Bodyfelt J~D, Hiller M and Kottos T 2007 {\em Europhys. Lett.\/} {\bf 78} 50003

\bibitem{Hiller2009}
Hiller M, Kottos T and Geisel T 2009 {\em Phys. Rev. A\/} {\bf 79} 023621

\bibitem{Santos2010}
Santos L~F and Rigol M 2010 {\em Phys. Rev. E\/} {\bf 81} 036206

\bibitem{Oganesyan2007}
Oganesyan V and Huse D~A 2007 {\em Phys. Rev. B\/} {\bf 75} 155111

\bibitem{Cassidy2009}
Cassidy A~C, Mason D, Dunjko V and Olshanii M 2009 {\em Phys. Rev. Lett.\/}
  {\bf 102} 025302

\bibitem{Choy1980}
Choy T~C 1980 {\em Phys. Lett.\/} {\bf 80A} 49

\bibitem{Haldane1980}
Haldane F~D~M 1980 {\em Phys. Lett.\/} {\bf 80A} 281

\bibitem{Haldane1981}
Haldane F~D~M 1981 {\em Phys. Lett.\/} {\bf 81A} 575

\bibitem{Choy1982}
Choy T~C and Haldane F~D~M 1982 {\em Phys. Lett.\/} {\bf 90A} 83

\bibitem{Krauth1991}
Krauth W 1991 {\em Phys. Rev. B\/} {\bf 44} 9772--9775

\end{thebibliography}

\end{document}